\documentclass[a4paper,11pt]{article}
\usepackage[dvips]{graphicx}
\usepackage{epsfig}
\begin{document}
\Large
\begin{center}

\textbf{ \\ Optical Observations of the CBS HZ
Her=Her X-1}\\

\copyright{} 2009 г.   Sazonov A.N.

Sternberg Astronomical Institute, Universitetskii pr. 13, Moscow, 119992 Russia

\end{center}
\vspace{12pt}

\subsection*{Introduction }

~ The high accuracy and long time span of photoelectric
observations allow them to be used for multifactor analyses and
refining some of the "fine" {} photometric effects in the light
curves of close binary systems. The results obtained can be
subsequently interpreted in terms of the model of mass flow from
the optical component of the close binary systems onto the
accretion disk of the neutron star, which can explain
satisfactorily the irregularities of the gaseous flow, the ``hot
spot''{}, and the presence of splashes moving in individual
Keplerian trajectories about the outer parts of the accretion disk
of the neutron star Her~X-1.

 \subsection*{Photometric peculiarities of the accretion formation
of the neutron star  Her X-1 and its geometry in 1986-1990}

~ When analyzing the photometric data in during the 1986 season
(fig01(a-d)), one must bear in mind that the scatter of individual
measurements in the  B, and especially, in the  W filter at times
exceeds, on the average, the corresponding standard errors (up
to~$0^m.3 0^m.5$), most likely, due to the physical variability of
the optical star, as we pointed out in our earlier
papers~\cite{Sazonov1987}, ~\cite{Luyty1989},~\cite{Sazonov2006}.

~ During the 1988 season the accretion formation in the system
increased in size, resulting in an increase of the duration of the
secondary minimum  Min ІІ, contrary to the conclusion made by
Kiliachkov~\cite{Kiliachkov1994}. This discrepancy can be
explained by the small number of observational data points used by
the above authors compared to the data set that we use in this
paper, and the lack of W- (or U-) and R-band in the former
analysis.

~ The system exhibits certain light variations at various
precession phases, which can be linked to the observed x-ray
variations of the Her X-1 source and which appear to depend on the
spatial orientation of the tilted and geometrically warped disk of
the neutron star (NS) precessing with a period of  $P_3 \cong
34^d.875$. The optical light variations considered may also be due
to the increase or decrease of the pulsar beam size as suggested
by Bisnovatyi-Kogan~\cite{Bisnovatyi-Kogan1975} and periodic
variation of the accretion rate onto the NS. These light
variations may explain the 35-day activity cycle of  Her X-1. The
lack of synchronization between orbital rotation and the axial
rotation of the optical component in the CBS can also explain the
precession of the accretion disk (AD) of the NS.

~ Many models are known to explain the main property of the 35-day
cyclic variations of the system - the periodic disappearance of
the x-ray flux for a terrestrial observer combined with the almost
constant x-ray luminosity of the NS~\cite{Kahabka1987}. This
conclusion follows from the fact that the reflection effect
continues to be observed even during the periods when no x-ray
flux from the system is recorded on the Earth.

~ The observations made in 1986-1988 exhibit certain light
variations in the precession phase interval (0.90-0.10), which can
be linked to the observed x-ray variations of the  Her X-1 source
and which appear to depend  on the spatial orientation of the
tilted and geometrically warped disk of the NS precessing with a
period of  $P_3 \cong 34^d.875$.

~ I favor the model of mass flow from the optical component of the
CBS onto the AD of the NS, which satisfactorily explains the
irregularities in the gaseous flow ~\cite{Sazonov2006},
~\cite{Kiliachkov1994}, ~\cite{Sazonov1990}, ~\cite{Bisikalo1994},
~\cite{Sheffer1997}, the "hot line"{}, and individual
splashes~\cite{Bochkarev1987} moving in individual Keplerian
trajectories about the outer parts of the AD of the  Her X-1
neutron  star~\cite{ Bochkarev1989}, ~\cite{Bochkarev1986}.

~ This model explains the variety of my observational data. Note
also that the numerical results obtained in earlier papers
~\cite{Sazonov1990} complement the proposed mechanism. The
comments made in the above paper unambiguously indicate that the
intersection of the family of the nearest ballistic orbits of the
gaseous flow produce concentrations of flow particles, which move
along certain spiral trajectories toward the accretor of the CBS
with ever increasing velocities due to the gravitational
acceleration caused by the primary component and which we can
interpret as a complex structure --- a shock --- interacting with
the ambient matter --- i.e., with the accretion disk of the
compact object. As a result, if a star, like HZ Her, fills
completely its inner critical Roche lobe  (ICRL), most of the mass
flows through the $L_1$ point, falls onto the NS, and feeds the
AD.

~ However, gas may partly scatter beyond the ICRL of the А7-type
star and the pulsar, mostly in the vicinity of the orbital plane
of the CBS and at small heights in the $\pm z$ plane
~\cite{Sazonov1990}, ~\cite{Bisikalo1994}, ~\cite{Kiliachkov1988},
~\cite{Lubow1993}; it can also escape from the system via the
$L_2$ point, crossing the neighboring ballistic trajectories of
the particles of the gaseous flow emerging from the optical
component of the system at small distances from the inner Lagrange
point $L_1$, i.e., before the jet encounters the outer edge of the
AD of the neutron star.

~ The location of the region where trajectories intersect depends
significantly on the initial velocities of flow particles. With
increasing absolute value of initial velocities of flow particles
the crossing region shifts downstream  so that the jet matter
should collide with the accretion disk of the neutron star before
the trajectories intersect with each other.

~ The intersection of ballistic trajectories of flow particles in
HZ Her= Her X-1 may explain, among other factors, the formation of
irregularities in the gaseous jet of the CBS, which result in the
flicker of the "hot line"{} in the HZ Her= Her X-1 CBS, which
appears in the region where the gaseous jet collides with the
disk-like envelope as observed at optical wavelengths.

~ We thus conclude that:

- The 1986 observations exhibit a certain increase of the
accretion rate onto the neutron star (fig02(a-d));

- During the 1987 season optical outbursts were observed in the
0.15-0.25 and 0.75-0.85 orbital phase intervals.
 Short flux outbursts in the phase interval near 0.015 were recorded
in the  R,V,B, and W filters with the amplitudes of  $0^m.10$,
$0^m.15$, $0^m.25$, and $0^m.35$, respectively;

- During the 1988 observing season the system behaved somewhat
chaotically near Min ІІ, especially at UV wavelengths, and this
behavior should be interpreted as a spatial manifestation of an
optically thick warped accretion disk;

- The WBVR light curves of 1989 reflect the dynamics of the
behavior of the system. Most of the observations were made during
the "off"{} phase, however, some  observations were made during
the high state. The accretion formation exhibits a somewhat
asymmetric behavior with the sign reversed compared to the
previous season. A "sharp"{} minimum, like in the 1987 season,
appeared once near Min. Near Min II the light curve has a
"classic"{} flat form.

- During the same observing season the light curve exhibited a
photometric peculiarity in the orbital phase interval
$\varphi=0.815-0.875$ resembling a cooling gaseous condensation or
a  "blob" {} (a blob of high-temperarture plasma circulating at
the outer edge of the AD of the neutron star) ~\cite{Crosa1980},
~\cite{Kippenhann1979}.

- The WBRI light curves of the 1990 season are indicative of a
certain decrease and equalization of the luminosities of both
halves of the accretion formations, and of a certain change of the
asymmetry of the light curve (especially in the B and V filters).
The size of the emitting region of the accretion formation appears
to have remained unchanged compared to its appearance during the
1988-1989 seasons.

 \subsection*{Dynamics of the Flux Variations in Min I and Min II
 and Mass Flow from the Optical Component}

~ Crosal~\cite{Crosa1980} performed the most detailed analysis of
a phenomenological model for the CBS HZ Her=Her X-1 assuming
constant mass flow from the primary star onto the NS. This model
explains the observational manifestations of the system that were
associated with the physical state of its  x-ray flux and that of
its accretion disk (AD).

~ A change in the mode of mass flow from the optical component of
the CBS to the accretion disk of the neutron star results in a
change of the geometry, degree of disk warp and, consequently, of
the disk temperature (from 18000 К to 25000 К ~\cite{Sazonov2006},
~\cite{ Kiliachkov1994}, ), resulting in the variations of
different duration (from 30-40 minutes to 2-3 hours) and amplitude
(from ~$0^m.2\pm0^m.3$ in the  R and V filters and up to
~$0^m.4\pm0^m.5$ in the B and W filters) appearing in the light
curve. This behavior shows up in an appreciable scatter of the
(W-B) color index compared to that of (B-V) and (V-R).

~ The variation of (W-B) (the cause) by up to ~$0^m.6$ (e.g.,
during the 1987 and 1992 seasons according to my observations) at
the same phases of the light curve results in  (the consequence)
changes in the geometric size and temperature of the "hot spot" {}
and accretion formations (fig03(a-b)).

~ Near Min I of the orbital period changes are observed in the
asymmetry of the accretion formation, which are associated with
different phases of the 35-day cycle. The "hot spot"{} exhibits a
certain evolution with respect to the central meridian of the
system even during the first cycle of the 35-day period. The
luminosity of the "hot spot" {} and its geometric size remained
more or less the same as in 1988. We see variations of the (W-B),
(B-V), (V-R), and (B-R) color indices as a function of the orbital
phase and for various precession phases.

~ During that observing year strong variations of the (W-B) color
index were observed, which are indicative of a certain decrease of
the temperature of the accretion formation in the "off" {} state
of the x-ray source, whereas the temperature and size of the
accretion formation remained within the traditional intervals of
the previous observing seasons. In 1990 the "hot spot" {}
exhibited appreciable evolution in the  "on" {} state relative to
the central meridian of the system.

~ We also observe in 1986-1998 a certain UV flux deficit during
the optical minima of  HZ Her.

~ Gaseous formations and the corona  scatter the radiation of the
optical component of the CBS, which is also observed during the
phases of optical eclipse ($\varphi _{orb.} = 0.97 - 0.03$), and
this fact should be taken into accounts in precision observations
and subsequent interpretation of the data obtained at these
phases.

~ The fact that W-band flux shows very strong variability
(amounting to $0^m.20 - 0^m.35$ over a single observing night
($\sim 3-4$ hours)) leads us naturally to conclude that we are
observing hot gas located in  a rather close vicinity of the
optical component of the CBS and flowing away from it with
velocities of about several hundred km/s ~\cite{Sazonov1990},
~\cite{Bochkarev1988}, namely, more than 210-260 km/s
~\cite{Sazonov1990} before encountering the accretor
--- the fact that was convincingly confirmed in later papers
~\cite{Bisikalo1995}.

~ Here we observe gaseous jets in the CBS, which undoubtedly
expand when reaching the AD of the neutron star, and these
processes must depend on the ambient properties  in the vicinity
of the CBS (and, in particular, in the vicinity of the accretion
disk of the neutron star) through which the jets move. This medium
is far from uniform.

\newpage
\begin{figure}[b]
\centerline{\epsfig{file=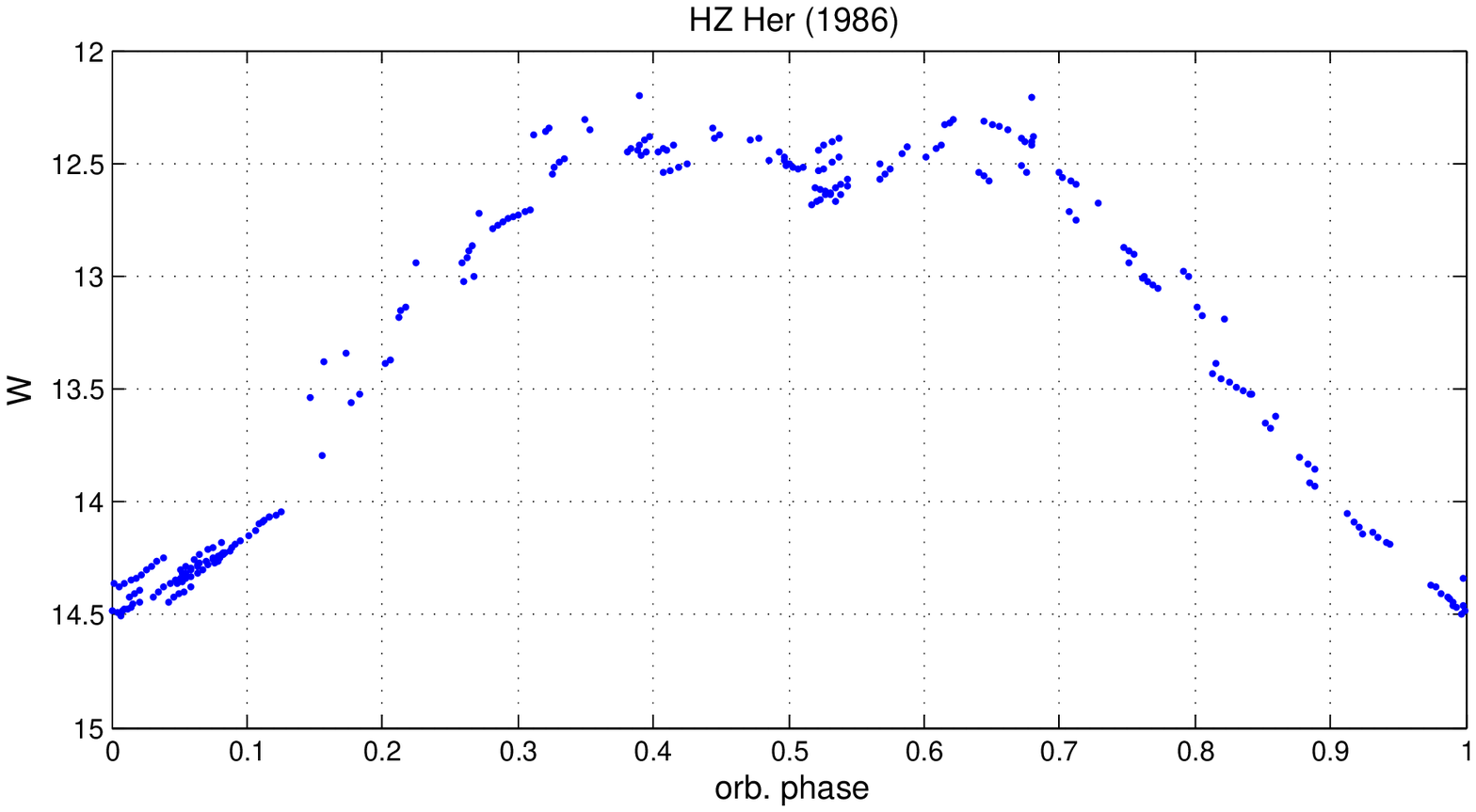,width=150mm}}
\caption{The 1986 W-band light curve folded with the period
$P=1^d.70016773$}
\end{figure}

\begin{figure}[b]
\centerline{\epsfig{file=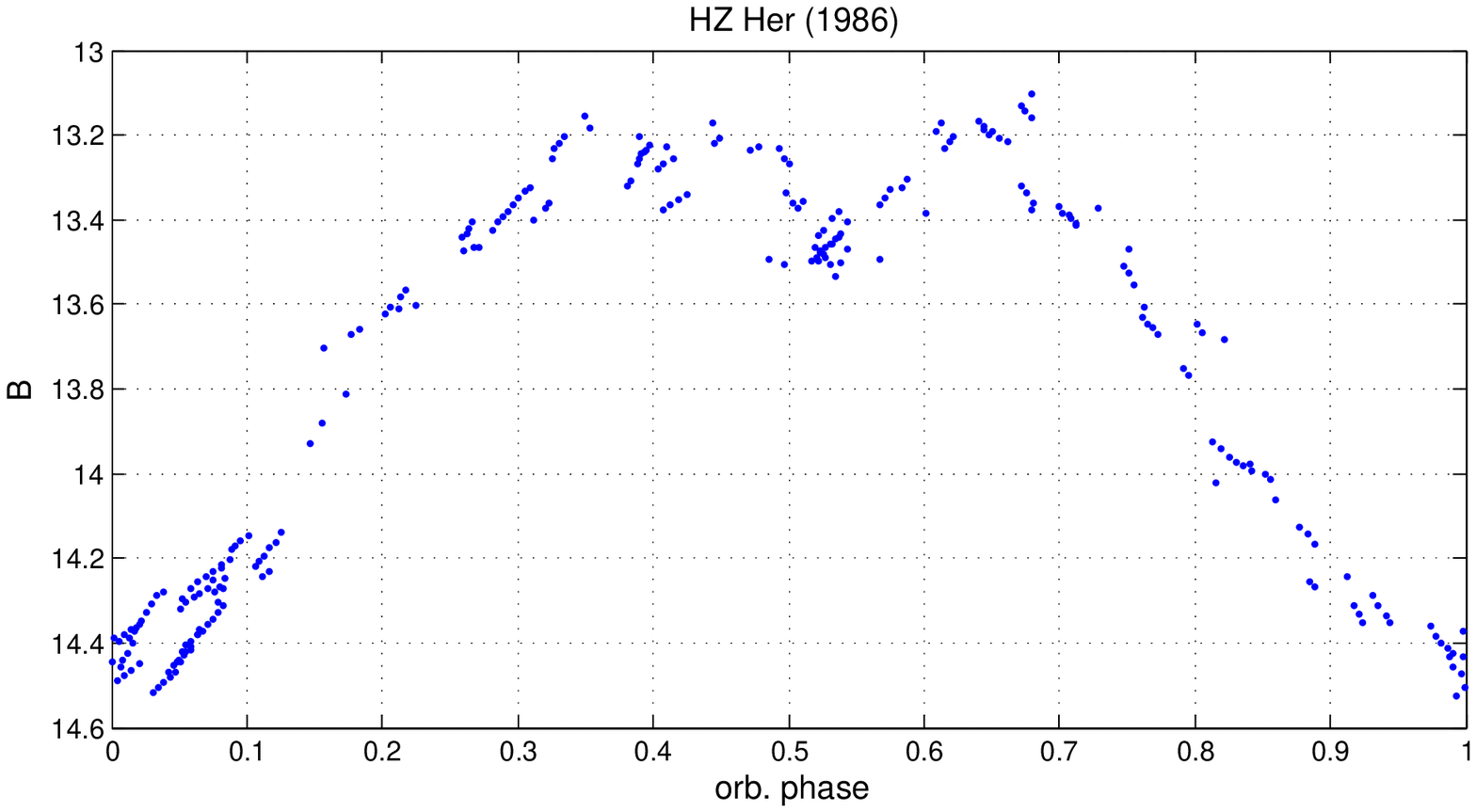,width=150mm}}
\caption{The 1986 B-band light curve folded with the period
$P=1^d,70016773$}
\end{figure}

\begin{figure}[b]
\centerline{\epsfig{file=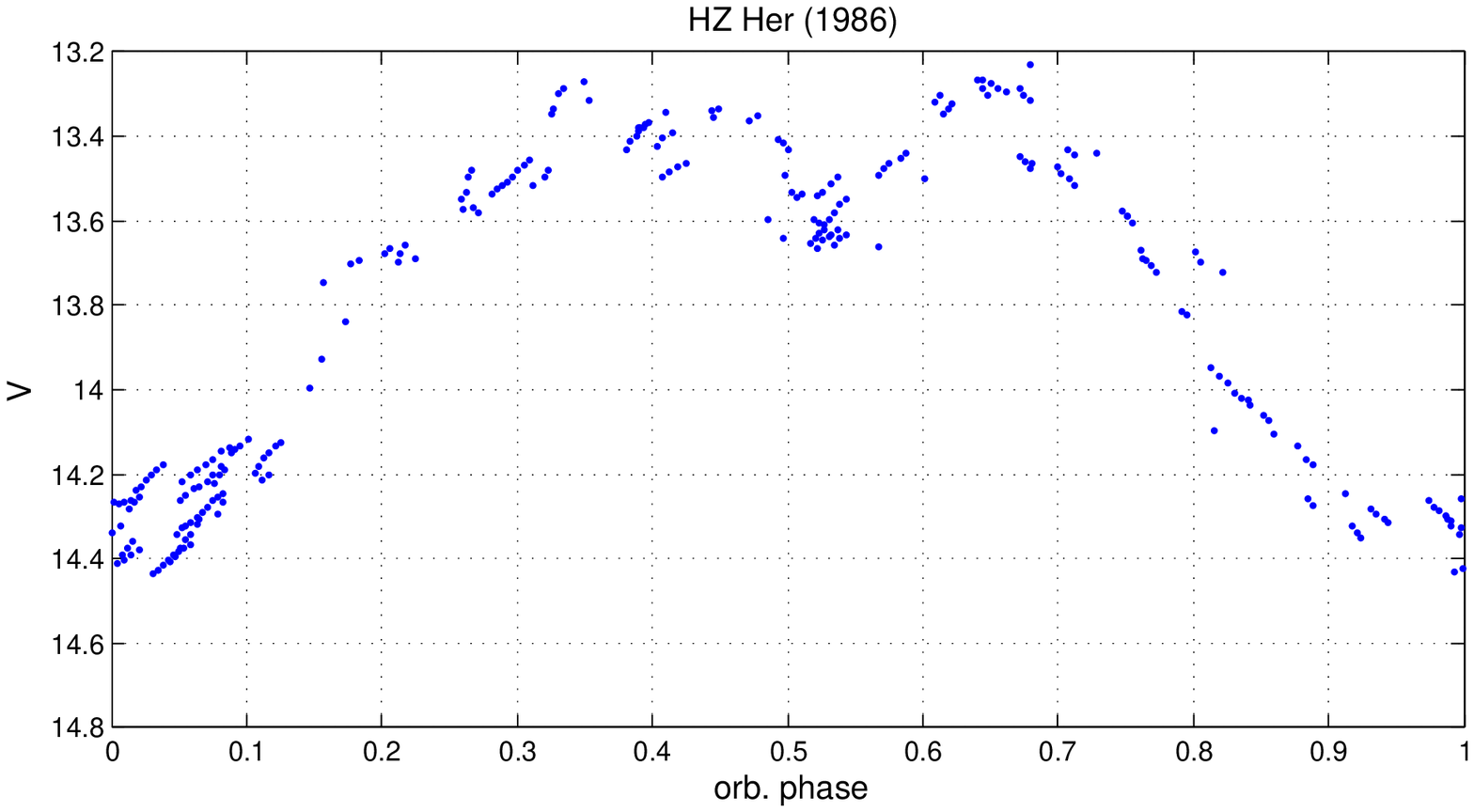,width=150mm}}
\caption{The 1986 V-band light curve folded with the period
$P=1^d,70016773$}
\end{figure}

\newpage
\begin{figure}[b]
\centerline{\epsfig{file=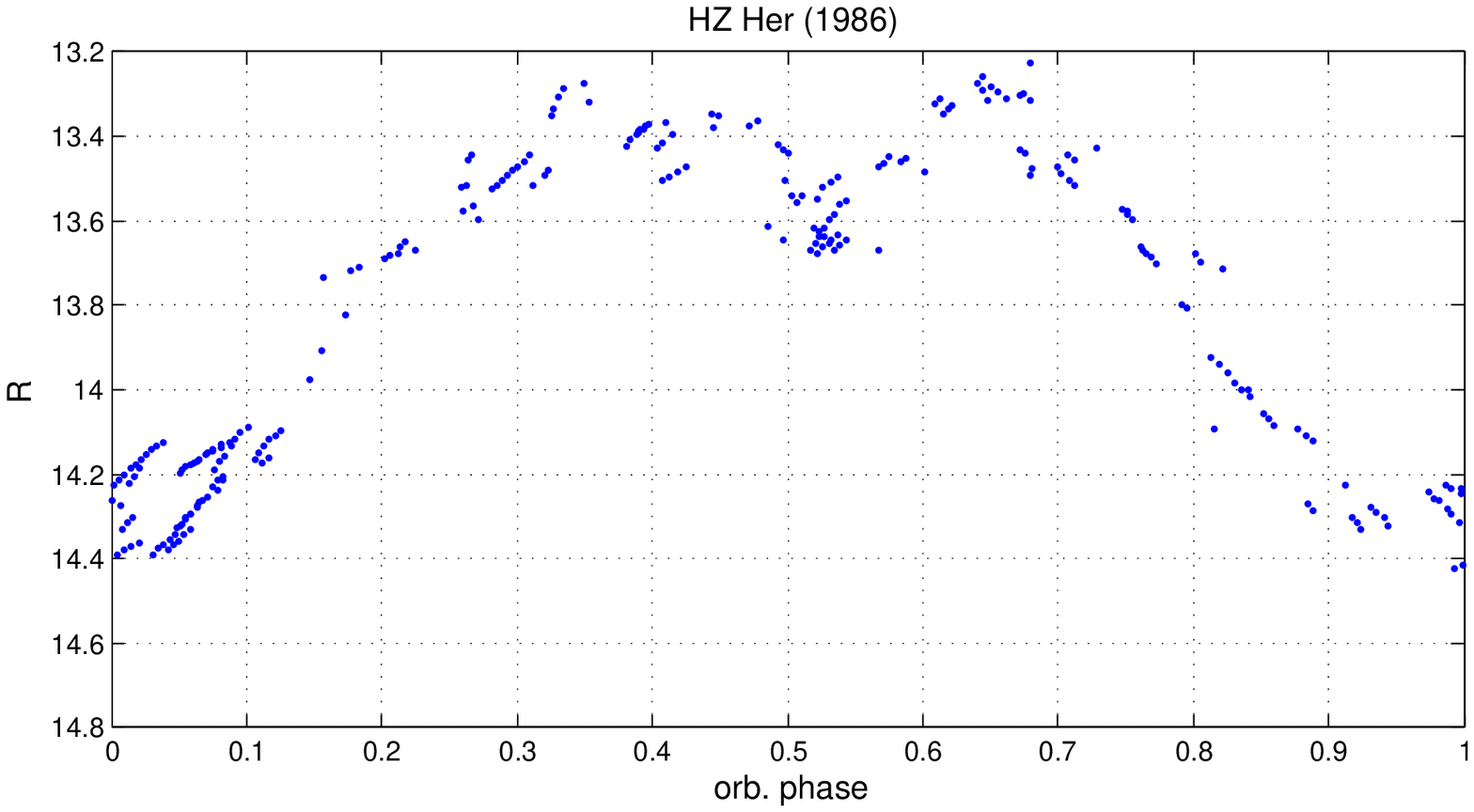,width=150mm}}
\caption{The 1986 R-band light curve folded with the period
$P=1^d,70016773$}
\end{figure}

\newpage
\begin{figure}[b]
\centerline{\epsfig{file=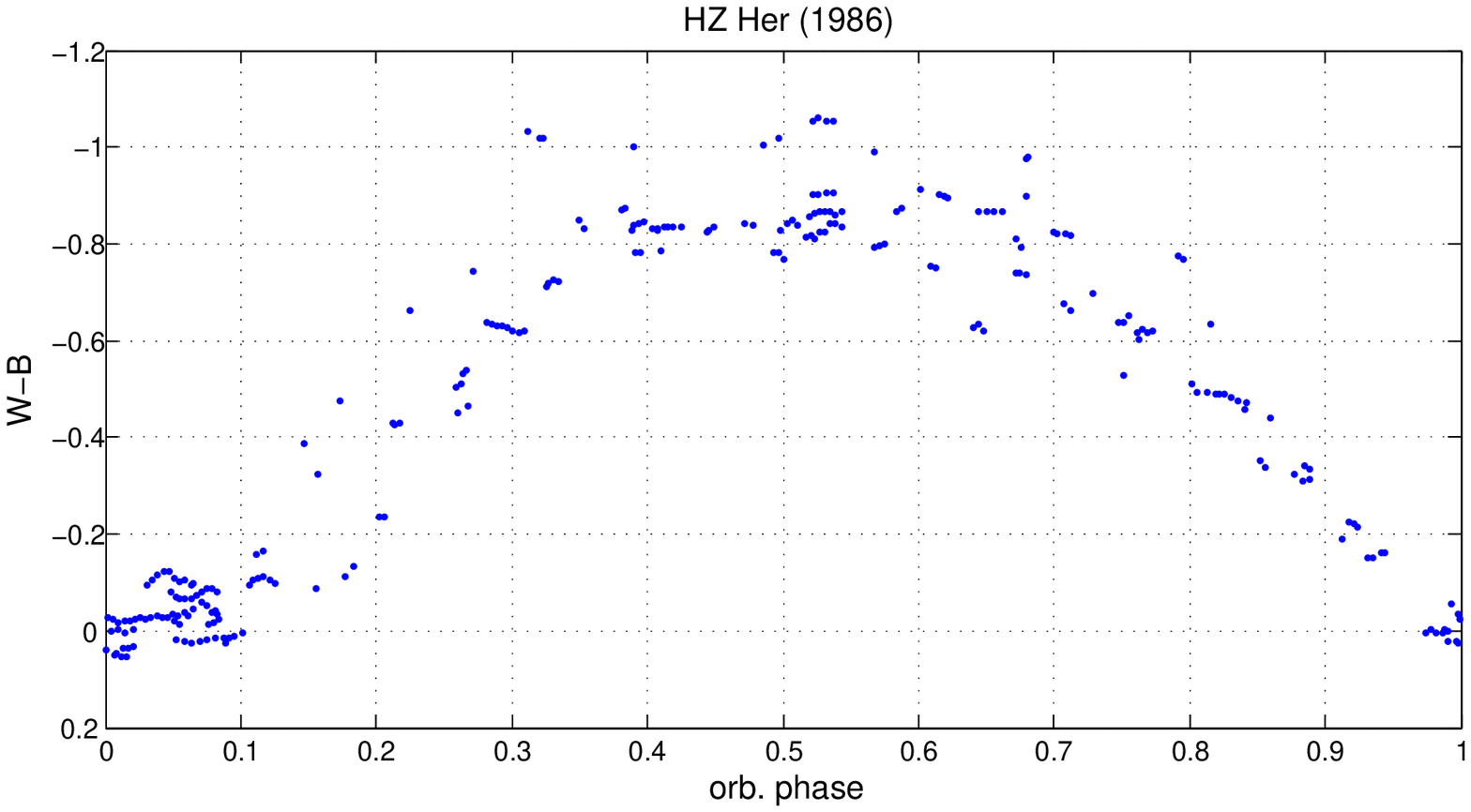,width=150mm}}
\caption{W-B color of HZ Her as a function of orbital phase
$\varphi$ for the 1986 data}
\end{figure}

\newpage
\begin{figure}[b]
\centerline{\epsfig{file=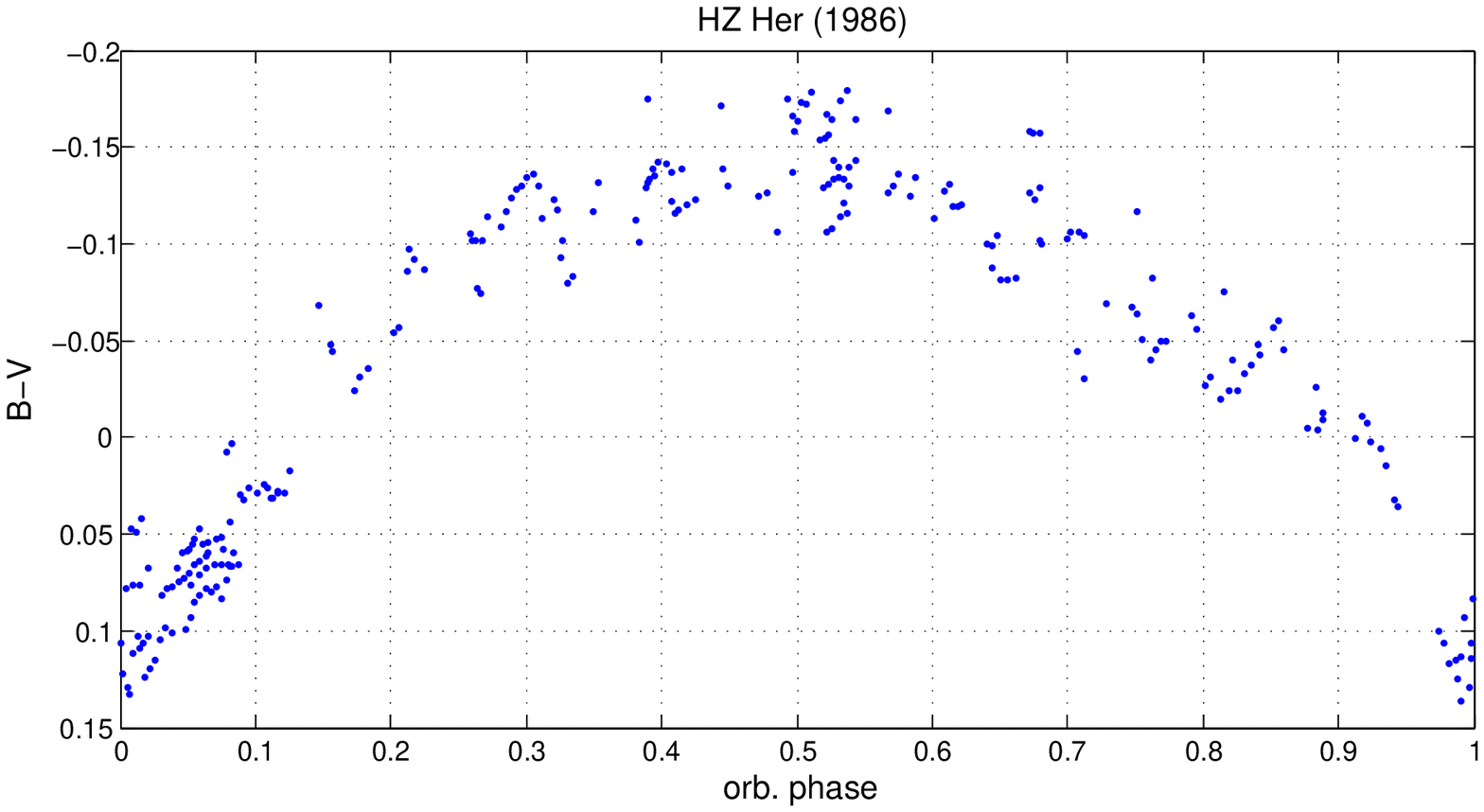,width=150mm}}
\caption{(B-V) color of HZ Her as a function of orbital phase
$\varphi$ for the 1986 data}
\end{figure}

\newpage
\begin{figure}[b]
\centerline{\epsfig{file=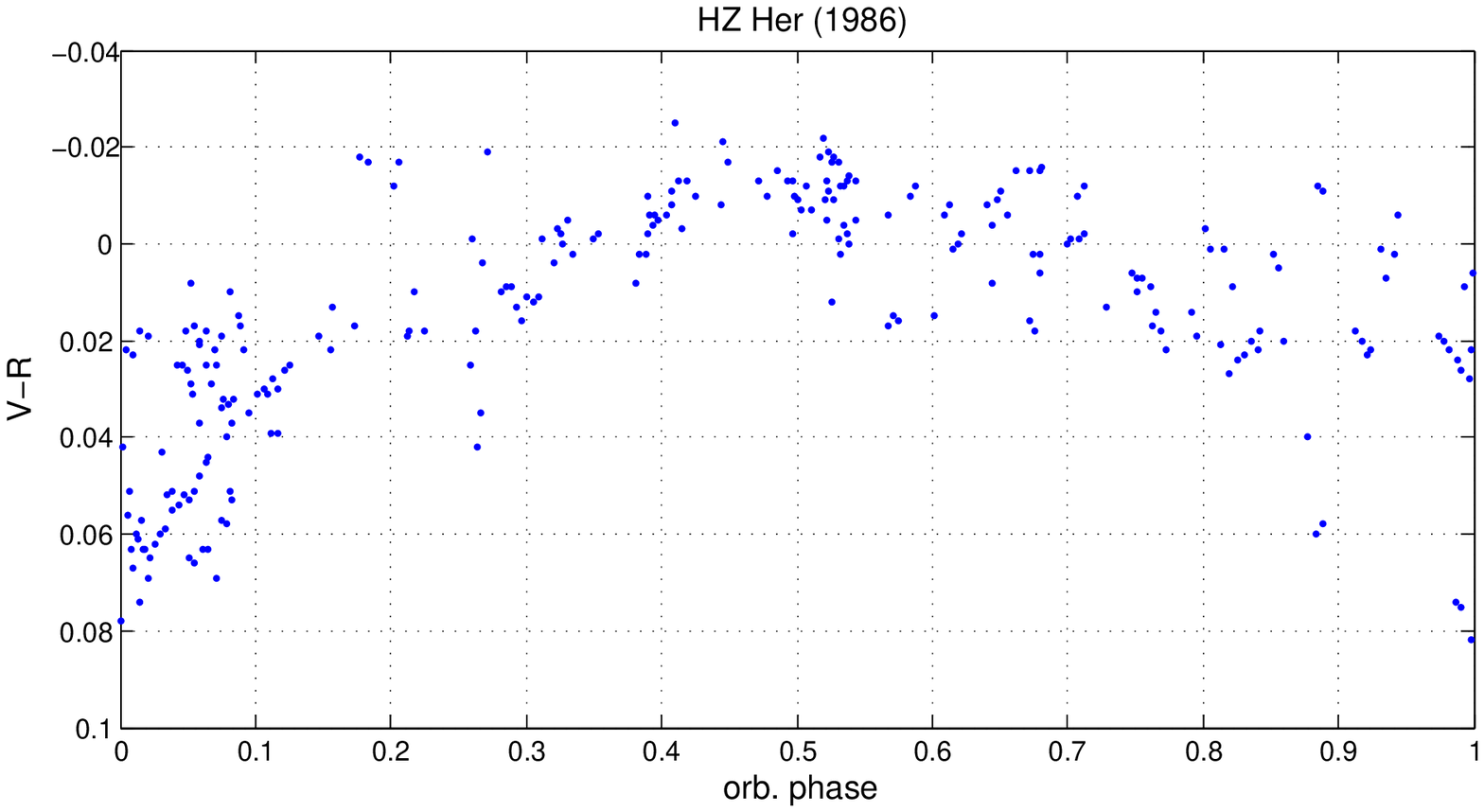,width=150mm}}
\caption{(V-R) color of HZ Her as a function of orbital phase
$\varphi$ for the 1986 data}
\end{figure}

\newpage
\begin{figure}[b]
\centerline{\epsfig{file=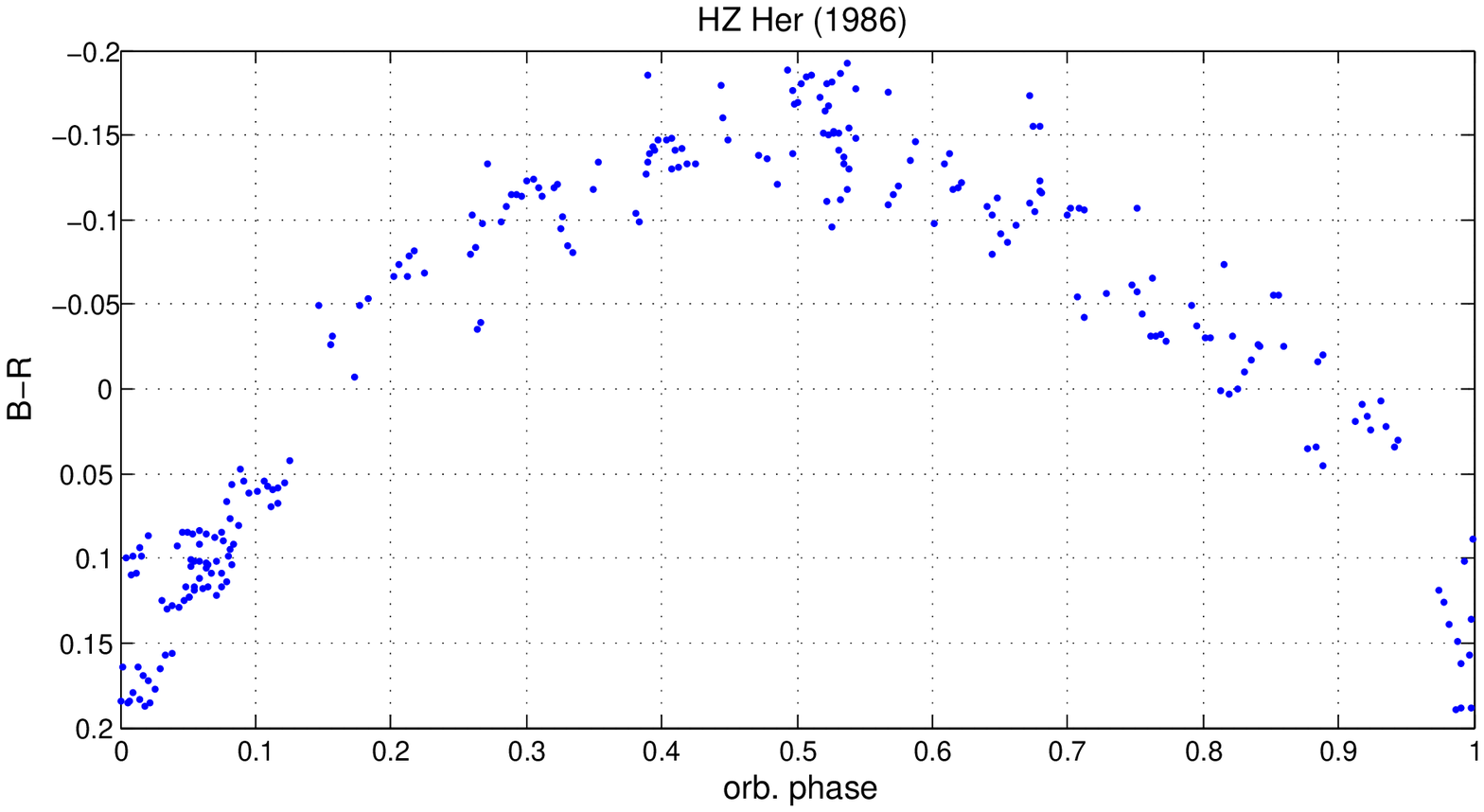,width=150mm}}
\caption{(B-R) color of HZ Her as a function of orbital phase
$\varphi$ for the 1986 data}
\end{figure}

\newpage
\begin{figure}[b]
\centerline{\epsfig{file=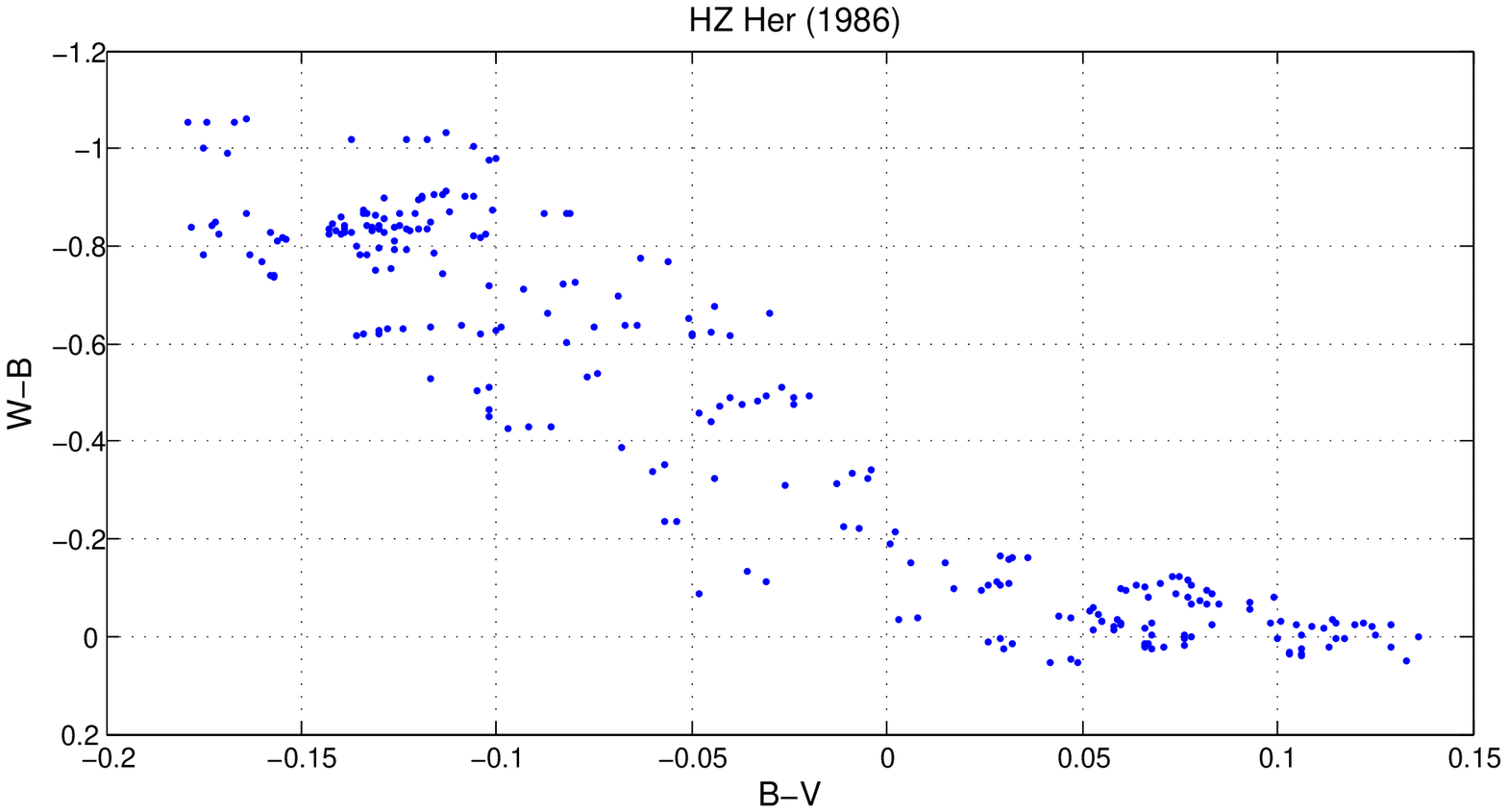,width=150mm}}
\caption{The 1986 (B-V)-(W-B) two-color diagram of HZ Her}
\end{figure}

\newpage
\begin{figure}[b]
\centerline{\epsfig{file=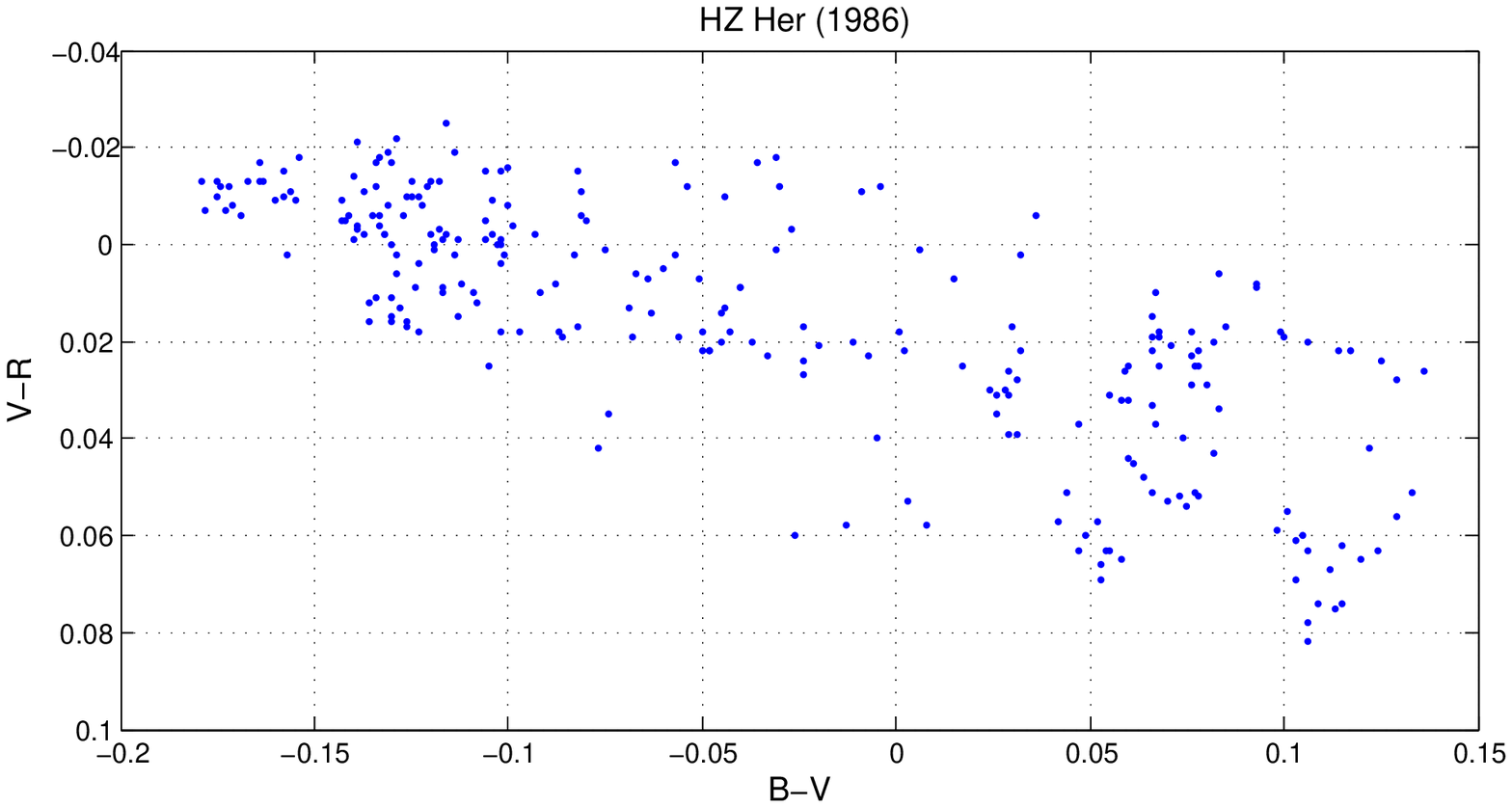,width=150mm}}
\caption{The 1986 (B-V)-(V-R) two-color diagram of HZ Her}
\end{figure}

\end{document}